\documentclass{article}

\usepackage[final, nonatbib]{neurips_2023_ml4ps}
\usepackage[utf8]{inputenc} 
\usepackage[T1]{fontenc}    
\usepackage{hyperref}       
\usepackage{url}            
\usepackage{booktabs}       
\usepackage{amsfonts}       
\usepackage{nicefrac}       
\usepackage{microtype}      
\usepackage{xcolor}         
\usepackage{siunitx}
\usepackage{graphicx}  
\usepackage{enumitem}

\title{Virtual EVE: a Deep Learning Model for Solar Irradiance Prediction}

\author{%
  Manuel Indaco \\
  Auburn University\\
  Auburn, AL 36849 \\
  USA\\
  \texttt{mzi0022@auburn.edu} \\
\And
  Daniel Gass \\
  University of Central Lancashire \\
  Preston, PR1 2HE \\
  United Kingdom\\
  \texttt{danielgass192@gmail.com} \\
\And
  William James Fawcett \\
  University of Cambridge \\
  Cambridge,  CB2 1TN \\
  United Kingdom\\
  \texttt{wjf29@cam.ac.uk} \\
\And
  Richard Galvez \\
  DataTalk AI\\
  New York, NY \\
  USA\\
 \texttt{richardagalvez@gmail.com} \\
\And
  Paul J. Wright \\
  Dublin Institute for Advanced Studies \\
  Dublin, D02 XF86 \\
  Ireland\\
 \texttt{paul.wright@dias.ie} \\
\And
  Andrés Muñoz-Jaramillo \\
  Southwest Research Institute \\
  Boulder, CO 80302 \\
  USA\\
 \texttt{amunozj@boulder.swri.edu} \\
}

\newcommand{\resolution}[1]{#1 × #1}

\begin{document}

\maketitle

\begin{abstract}
Understanding space weather is vital for the protection of our terrestrial and space infrastructure. In order to predict space weather accurately, large amounts of data are required, particularly in the extreme ultraviolet (EUV) spectrum. An exquisite source of information for such data is provided by the Solar Dynamic Observatory (SDO), which has been gathering solar measurements for the past 13 years. However, after a malfunction in 2014 affecting the onboard Multiple EUV Grating Spectrograph A (MEGS-A) instrument, the scientific output in terms of EUV measurements has been significantly degraded. Building upon existing research, we propose to utilize deep learning for the virtualization of the defective instrument. Our architecture features a linear component and a convolutional neural network (CNN) -- with EfficientNet as a backbone. The architecture utilizes as input grayscale images of the Sun at multiple frequencies -- provided by the Atmospheric Imaging Assembly (AIA) -- as well as solar magnetograms produced by the Helioseismic and Magnetic Imager (HMI). Our findings highlight how AIA data are all that is needed for accurate predictions of solar irradiance. Additionally, our model constitutes an improvement with respect to the state-of-the-art in the field, further promoting the idea of deep learning as a viable option for the virtualization of scientific instruments.
\end{abstract}

\section{Introduction}
The behavior of Earth's atmosphere and thermosphere is principally governed by the extreme ultraviolet (EUV) radiation emitted by the Sun \cite{Muller1987}. This solar radiation plays a pivotal role in affecting various aspects of both terrestrial and space-based infrastructure. Solar flares and geomagnetic storms interact with the ionosphere, leading to potential issues in long-range communications, or induce currents in terrestrial power lines, with the potential to lead to catastrophic power outages. Similarly, solar activity can induce electrostatic discharges on orbiting satellites, leading to hardware degradation; notably phenomena such as single-event upsets can cause data corruption or interfere with the power distribution system. As a consequence, a thorough understanding of solar EUV radiation is crucial for the protection of this critical infrastructure. 

Historically, many missions have been designed and launched for the study of our star. One such mission is NASA's Solar Dynamics Observatory (SDO) \cite{Pesnell2012}, launched in 2010 as a part of the ``Living With a Star" program \cite{livingstar}. The SDO spacecraft features three main instruments: the Atmospheric Imaging Assembly (AIA) \cite{Lemen2012}, which captures full-disk images of the Sun at \resolution{4096} resolution in seven EUV channels, two UV channels, and one visible channel, the Helioseismic and Magnetic Imager (HMI) \cite{Scherrer2012}, which delivers photospheric vector magnetograms and Dopplergrams of the full Sun at \resolution{4096} resolution, and the EUV Variability Experiment (EVE) \cite{Woods2012}, composed by the Multiple EUV Grating Spectrograph A (MEGS-A) and MEGS-B modules, which provide measurements of the solar spectrum irradiance over 39 spectral lines (or ions). Unfortunately, a capacitor short in 2014 led to the loss of MEGS-A, substantially diminishing the scientific return on EUV data. To fill this gap, Szenicer et al.~\cite{Szenicer2019} proposed a deep learning model for the virtualization of the compromised instrument, demonstrating encouraging results in the reconstruction of the solar irradiance over 14 ions channels by utilizing exclusively 8 AIA channels. In this investigation, we build upon the work presented in \cite{Szenicer2019} and \cite{Tremblay2023Enhancing}, by introducing multiple elements of novelty: 1) we build a model that not only takes as input 9 AIA channels, but also ingests HMI magnetograms resolved into $x,y,z$ components; 2) we provide empirical evidence that AIA alone is enough for accurately predicting solar irradiance; 3) we train our model to predict all ions probed by both MEGS-A and MEGS-B; and 4) we improve upon the model's accuracy.

\section{Dataset}
For the training of our model, we consider data produced by the AIA, HMI, and EVE instruments between 2010 and 2014. The choice of not utilizing data beyond 2014 for training purposes is driven by the fact that our model is trained in a supervised fashion; as such, we are limited by the operational lifetime of MEGS-A. We source our data from the SDOMLv2 dataset\footnote{\url{https://www.sdoml.org}}. This dataset includes AIA and HMI images for the time range 2010 to 2020, as well as EVE measurements from 2010 to 2014, conveniently stored in zarr format. In the following, a brief description of each individual data product is provided. More details in \cite{Galvez2019}.
\begin{description}
\item[AIA] The AIA dataset consists of \resolution{512} pixel images at a 6-minute cadence calibrated to Level 1.5 \cite{Lemen2012}. All images in the SDOML dataset have had the following applied: images are corrected for the degradation of the detector \& normalization for exposure time,  re-alignment so the $x$ and $y$ axes align with the solar west and north respectively, re-scaling of the solar disk so that occupies the same angular size in all images. We consider the wavelengths  [94, 131, 171, 193, 211, 304, 335, 1600, 1700] $\si{\angstrom}$, excluding 4500 $\si{\angstrom}$ as it provides no valuable information for EUV irradiance prediction.
\item[HMI] The HMI images are also in \resolution{512} resolution but with a 12-minute cadence. The magnetic field has been resolved into  $B_x$, $B_y$ and $B_z$ components. Like with the AIA data the HMI images are spatially aligned, so that the solar disk occupies the same area and axes are aligned with the solar north and west.  
\item[EVE] The  EVE measurements in the dataset cover a time range between May 2010 and May 2014. The observations are calibrated and provided with physical units Wm$^{-2}$, scaled to {1 astronomical} unit and corrected for degradation. Temporal synchronization is applied to match the AIA data, hence resulting in available EVE measurements at a 6-minute cadence.
\end{description}

\begin{figure}[!ht]
\centering
{\includegraphics[width=0.95\linewidth]{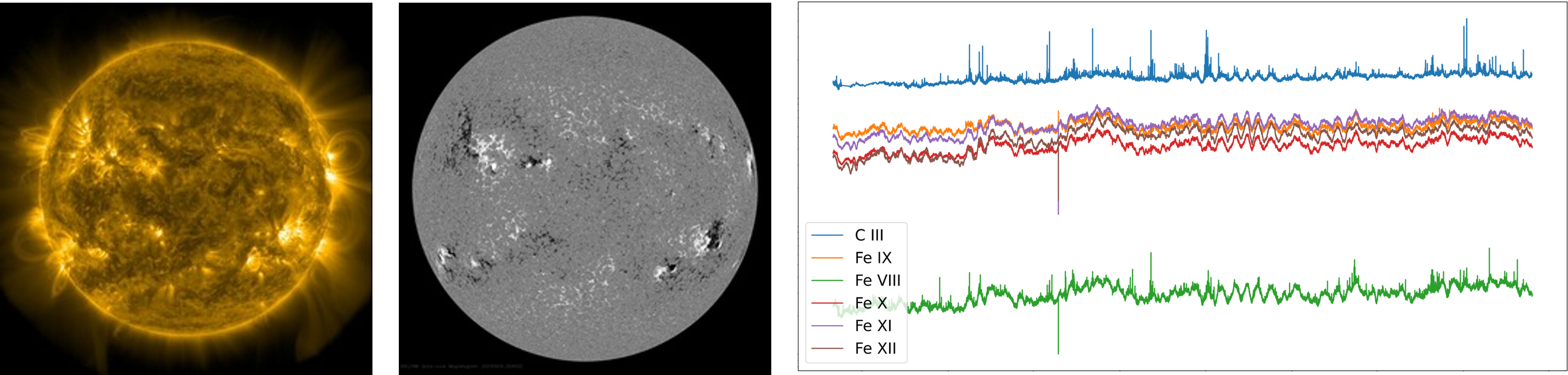}}
\caption{Example data for (left) AIA, (center) HMI and (right) EVE.}
\label{fig: data_sample}
\end{figure}

\section{Methodology}
Following the steps taken in \cite{Szenicer2019}, and building upon the model developed by \cite{Tremblay2023Enhancing}, we leverage deep learning to create a virtual EVE instrument. We augment the input dimension by adding a channel for AIA images (for a total of nine), and introducing HMI images for the three magnetic field components. Additionally, we increase the number of predicted ions from the original total of 14, which covers solely MEGS-A, to 38, hence including MEGS-B observations. 
This has the important consequence in that model predictions for MEGS-B can be compared to real observations up to the present, increasing confidence in the related MEGS-A predictions.
The architecture consists of a combination of a linear and a convolutional component (EfficientNet \cite{Tan2019} used as a backbone), which we refer to as a "hybrid model". The output of the network is generated by combining the outputs of the individual network components as follows:
\begin{equation}
    O_{\mathrm{total}} = O_{\mathrm{linear}} + \lambda O_{\mathrm{CNN}}
\end{equation}
with $\lambda$ being a weight factor tuning the impact of the CNN on the prediction of the network, and $O_{X}$ being a network component output. As the input-output relations can be well captured through a simple linear regression, $\lambda$ is assigned a small value, thus effectively having the CNN act as a corrector where the linear network struggles to capture more complex, non-linear relations. The linear component is a single-layer feedforward network preceded by a dropout layer. The choice of introducing the linear layer is justified by the fact that by performing a linear combination of the input pixel images, the output is expected to be proportional to the total irradiance, at minimum when MEGS-A channels overlap with AIA wavelength. We split the network training into two halves: in the first half solely the linear component is trained, while the CNN weights are frozen. Furthermore, during the training of the linear layer, the $\lambda$ factor is set to 0, resulting in an irradiance prediction based on solely the linear component alone. After a predetermined number of epochs, the CNN block is activated, while the weights belonging to the linear component are frozen. Huber Loss is chosen for training the architecture, which reduces the sensitivity to outliers with respect to a more traditional MSE loss.

\section{Results}

\subsection{Simulations Setup}
We conducted an extensive testing campaign to advance the development of our virtual EVE instruments. In particular, experiments encompassed both general simulation parameters and neural network architecture hyperparameters and training. Regarding simulations, we systematically explored two crucial aspects: 1) the cadence of input data, ranging from intervals of tens of minutes to hours, and 2) the source of input data, i.e., either AIA or HMI data individually, or a fusion of both sources. Furthermore, we explored the significance of specific input channels' on prediction quality, aiming to avoid the introduction of noise that could adversely affect network output. In all scenarios, we applied normalization to both input and output data, employing traditional mean subtraction and division by standard deviation. From a training and network parameter perspective, we experimented with various batch sizes and different learning rates within the range of 0.01 to 0.0001. Furthermore, we conducted informal tests on the EfficientNet backbone size (specifically employing the architectures b3, b5, and b7, which range from approximately 10 million to 70 million parameters) to evaluate the impact of network dimensions on irradiance prediction accuracy.

\subsection{Model Performance}
Overall, our experiments demonstrate how the linear component typically performed well in all situations, with the CNN providing an additional improvement when properly tuned. In fact, the sweep on the learning rate revealed how values above 0.001 resulted in a weight-explosion phenomenon in the CNN. Despite the weight factor always being set to a low value (here set to a value of 0.1, which we empirically determined during the tuning of the model), the CNN misbehavior dominated the output of the linear part, leading to an overall instability of the training. In merit of the backbone architecture, our tests demonstrate how larger models perform generally better, displaying a smoother behavior both in terms of global and local (i.e. referred to individual channels)  metrics.

Particularly interesting information derives from the observation of the effect of the input data source. Our experiments revealed how HMI-only input overall leads to the worst performance. In addition, in opposition to theoretical expectations, we discovered that the addition of HMI data to AIA data does not introduce an improvement. As a matter of fact, all the conducted analyses suggest  AIA data is the only information (considering the data chosen data sources) required to effectively predict solar irradiance. 

\begin{figure}[ht]
\centering
{\includegraphics[width=1\linewidth]{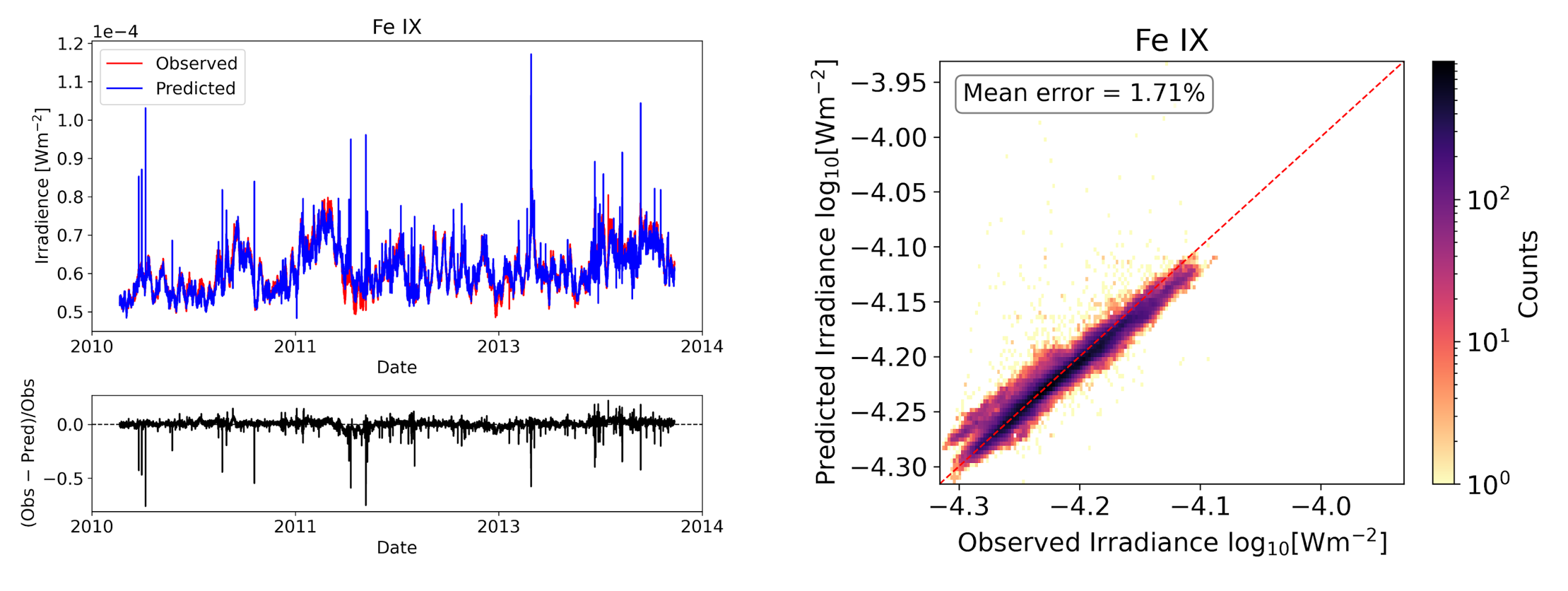}}
\caption{Irradiance observations and predictions for Fe IX (wavelength 171.1\,$\si{\angstrom}$). (top left) unnormalized observed and predicted irradiance, and (bottom left) residuals between predicted and real data over time between 2010 and 2014. (Right) summary of predicted vs. observed values distribution.}
\label{fig: FeIX}
\end{figure}

\begin{figure}[ht]
\centering
{\includegraphics[width=1\linewidth]{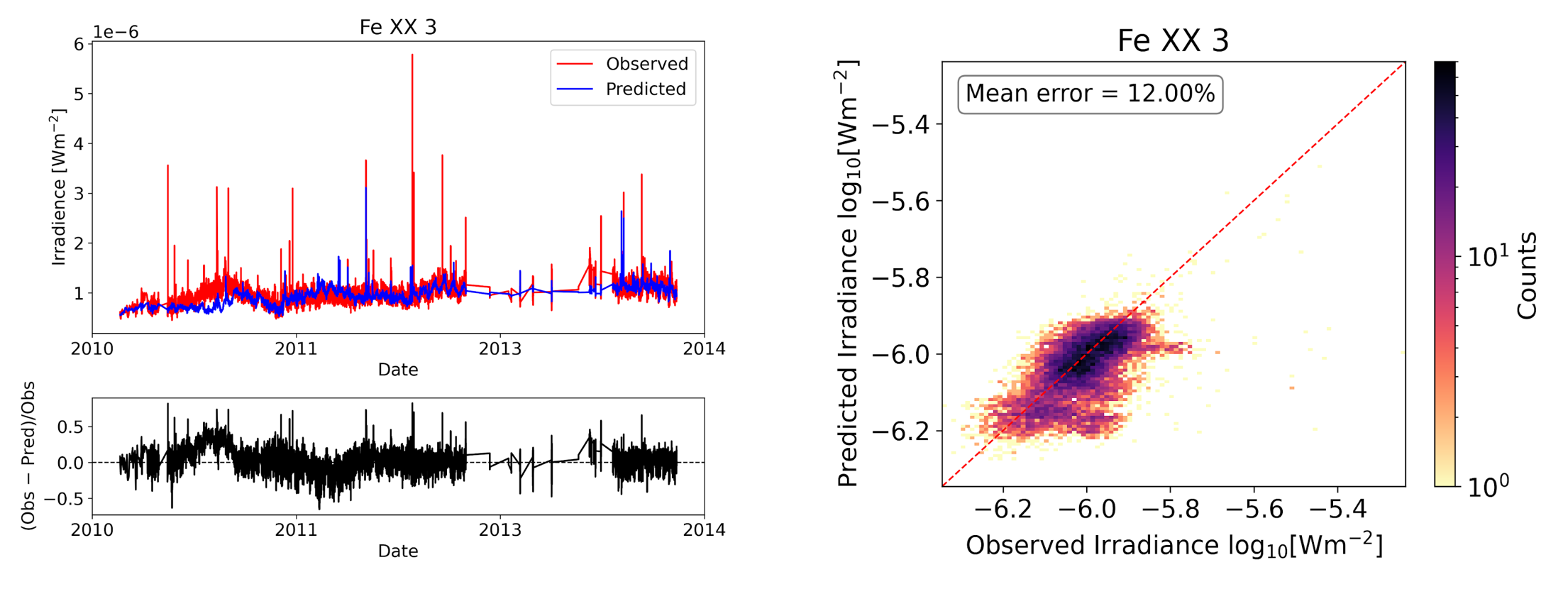}}
\caption{
Irradiance observations and predictions for Fe XX 3 (wavelength 721.61\,$\si{\angstrom}$). (top left) unnormalized observed and predicted irradiance, and (bottom left) residuals between predicted and real data over time between 2010 and 2014. (Right) summary of predicted vs. observed values distribution.
}
\label{fig: FeXX3}
\end{figure}

While these observations apply at the global level, further insight can be gained by looking at specific channels. In particular, we observed how the network performs extremely well (Fig.~\ref{fig: FeIX}) for ion channels whose wavelength is covered by AIA frequencies. Conversely, for ions whose frequency is far from any AIA channel, the network demonstrates little to no learning capability (Fig.~\ref{fig: FeXX3}). This is reasonable considering AIA data are the main driver of network performance.


\section{Conclusions}
The SDO mission provides invaluable scientific data for the understanding of our star, which is crucial for the protection of both terrestrial and space-based infrastructure. Unfortunately, as the MEGS-A instrument onboard SDO suffered from a failure in 2014, the scientific throughput in terms of EUV data has been significantly reduced since that date. 

To overcome this problem, we have built a composite deep learning model, inspired by \cite{Szenicer2019} and \cite{Tremblay2023Enhancing}, to virtualize the EVE instrument. We have introduced several improvements to previous work, including producing MEGS-B irradiances, and the ability to use HMI magnetograms as part of the input data. We have also performed a detailed analysis of the effect of input cadence on improving performance.

Though a direct comparison with the reference model \cite{Szenicer2019} is not possible (as our results summary includes both flaring and non-flaring data together whereas \cite{Szenicer2019} only includes non-flaring data), our model achieves similar levels of accuracy on several ion lines to \cite{Szenicer2019}, and outperforms the physics-based model \cite{wright_2019_2587015}. 

Furthermore, previous models could only predict MEGS-A ions whereas our model also predicts the related MEGS-B ions, substantially enhancing the capabilities of the virtual instrument. As MEGS-B is still operational, cross-checking of the model performance with live data can be performed. Although it is impossible to get real MEGS-A data for the present, if the model predictions are good for MEGS-B ions then it provides some confidence that MEGS-A predictions are also accurate. 
 
Finally, in a novel addition, we tested the introduction of HMI data to the model. This was because HMI data may have improved the model accuracy during flaring times, where the quality of AIA data degrades or is completely absent. 
Notably, we discovered how the injection of HMI data results in little to no improvements in the prediction capability, demonstrating how AIA is all you need for successful solar EUV irradiance reconstruction.

\section*{Acknowledgement}
This work is the research product of FDL-X (\url{www.fdlxhelio.org}), an initiative of FDL.AI (FDL.ai). FDL.AI is a public/private partnership between NASA, Trillium Technologies (\url{https://trillium.tech/}) and commercial AI partners Google Cloud and Nvidia, developing open science for all Humankind. We thank the 4pi team of 2022 for giving us access to their code, which was used as an initial starting point for this work.

\medskip

\bibliographystyle{unsrt}
\bibliography{references}

\begin{thebibliography}{10}

\bibitem{Muller1987}
C.~Muller.
\newblock Theory of planetary atmospheres: An introduction to their physics and
  chemistry.
\newblock {\em Eos, Transactions American Geophysical Union},
  68(51):1795--1795, December 1987.
\newblock DOI: \url{ https://doi.org/10.1029/EO068i051p01795}.

\bibitem{Pesnell2012}
W.~Dean Pesnell, B.~J. Thompson, and P.~C. Chamberlin.
\newblock {The Solar Dynamics Observatory (SDO)}.
\newblock {\em Solar Physics}, 275(1-2):3--15, January 2012.
\newblock DOI: \url{https://doi.org/10.1007/s11207-011-9841-3}.

\bibitem{livingstar}
G.~L. Withbroe.
\newblock {\em Living with a star}, pages 45--51.
\newblock American Geophysical Union, 2013.

\bibitem{Lemen2012}
J.~R.~Lemen et~al.
\newblock {The Atmospheric Imaging Assembly (AIA) on the Solar Dynamics
  Observatory (SDO)}.
\newblock {\em Solar Physics}, 275(1-2):17--40, January 2012.
\newblock DOI: \url{https://doi.org/10.1007/s11207-011-9776-8}.

\bibitem{Scherrer2012}
P.~R.~Scherrer et~al.
\newblock {The Helioseismic and Magnetic Imager (HMI) Investigation for the
  Solar Dynamics Observatory (SDO)}.
\newblock {\em Solar Physics}, 275(1-2):17--40, January 2012.
\newblock DOI: \url{https://doi.org/10.1007/s11207-011-9834-2}.

\bibitem{Woods2012}
T.~N.~Woods et~al.
\newblock {Extreme Ultraviolet Variability Experiment (EVE) on the Solar
  Dynamics Observatory (SDO): Overview of Science Objectives, Instrument
  Design, Data Products, and Model Developments}.
\newblock {\em Solar Physics}, 275(1-2):17--40, January 2012.
\newblock DOI: \url{https://doi.org/10.1007/s11207-009-9487-6}.

\bibitem{Szenicer2019}
A.~Szenicer et~al.
\newblock {A deep learning virtual instrument for monitoring extreme UV solar
  spectral irradiance}.
\newblock {\em Science Advances}, 5(10):1--10, October 2019.
\newblock DOI: \url{https://doi.org/10.1126/sciadv.aaw6548}.

\bibitem{Tremblay2023Enhancing}
B.~Tremblay et~al.
\newblock Enhancing {Observational} {Capabilities} of {EUV}-observing
  {Satellites} to {Estimate} {Spectral} {Irradiance}.
\newblock {\em Bulletin of the AAS}, 55(7), September 2023.
\newblock https://baas.aas.org/pub/2023n7i110p05.

\bibitem{Galvez2019}
R.~Galvez et~al.
\newblock {A Machine-learning Data Set Prepared from the NASA Solar Dynamics
  Observatory Mission}.
\newblock {\em The Astrophysical Journal Supplement Series}, 242(1):1--12, May
  2019.
\newblock \url{10.3847/1538-4365/ab1005}.

\bibitem{Tan2019}
M.~Tan and Q.~V. Le.
\newblock {EfficientNet: Rethinking Model Scaling for Convolutional Neural
  Networks}.
\newblock {\em Proceedings of the 36th International Conference on Machine
  Learning, Long Beach (CA, USA)}, pages 6105--6114, June 2019.

\bibitem{wright_2019_2587015}
P.~J.~Wright et~al.
\newblock {DeepEM: Demonstrating a Deep Learning Approach to DEM Inversion}.
\newblock March 2019.
\newblock DOI: \url{https://doi.org/10.5281/zenodo.2587015}.

\end{thebibliography}

\end{document}